\def\half{\frac{1}{2}}

\def\eq#1{Eq.\,(\ref{#1})}
\documentclass{iopart}\usepackage{cite}\usepackage{graphicx}
\usepackage{mathrsfs} 
\begin{document}
\hspace*{3.5 in}CUQM-143\\
\markboth{R.~L.~Hall \& P. Zorin}{Dirac eigenvalues}

\title{Dirac eigenvalues for a softcore Coulomb potential in $d$ dimensions}

\author{Richard~L.~Hall and Petr~Zorin}
\address{Department of Mathematics and Statistics, Concordia
University,\\ 1455 de Maisonneuve Boulevard West, Montr\'eal,
Qu\'ebec, Canada H3G 1M8\\ rhall@mathstat.concordia.ca,~petrzorin@yahoo.com}

\begin{abstract} A single fermion is bound by a softcore central Coulomb potential $V(r) = -{v}/{(r^q+b^q)^{\frac{1}{q}}}$ , $v>0,~b>0,~q \ge 1$, in $d>1$ spatial dimensions. Envelope theory is used to construct analytic lower bounds for the discrete Dirac energy spectrum. The results are compared to accurate eigenvalues obtained numerically.
\end{abstract}

\vskip 0.2in
\maketitle
\section{Introduction}
 For a central potential in $d$ dimensions the Dirac equation can be written \cite{jiang} in natural units $\hbar=c=1$ as
\begin{equation*}\label{eq1}
i{{\partial \Psi}\over{\partial t}} =H\Psi,\quad {\rm where}\quad  H=\sum_{s=1}^{d}{\alpha_{s}p_{s}} + m\beta+V,
\end{equation*}
$m$ is the mass of the particle, $V$ is a spherically symmetric vector potential, and $\{\alpha_{s}\}$ and $\beta$  are the Dirac matrices which satisfy anti-commutation relations; the identity matrix is implied after the potential $V$. For stationary states, some algebraic calculations in a suitable basis lead to a pair of first-order linear differential equations in two radial functions $\{\psi_1(r), \psi_2(r)\}$, where $r = ||\mathbf{r}||.$  For $d > 1,$ these functions vanish at $r = 0$, and, for bound states, they may be normalized by the relation 
\begin{equation*}\label{eq2}
(\psi_1,\psi_1) + (\psi_2,\psi_2) = \int\limits_0^{\infty}(\psi_1^2(r) + \psi_2^2(r))dr = 1.
\end{equation*}
We use inner products {\it without} the radial measure $r^{(d-1)}$ because the factor $r^{\frac{(d-1)}{2}}$ is already built into each radial function. It has been shown, for example by Jiang \cite{jiang} (using algebraic ladder-operator methods), that these functions satisfy the following coupled radial equations
\begin{eqnarray}
E\psi_1 &=& (V+m)\psi_1 + (-\partial + k_{d}/r)\psi_2\label{eq3},\\
E\psi_2 &=& (\partial + k_{d}/r)\psi_1 + (V-m)\psi_2\label{eq4},
\end{eqnarray}
where $\partial = \partial/\partial r,$ $k_1 = 0,$ $k_{d}=\tau(j+{{d-2}\over{2}}),~d >1$, $\tau = \pm 1$, and $j=1/2, 3/2, 5/2, \ldots$ is the total angular momentum quantum number. We note that the variable $\tau$ is sometimes written as $\omega$, as, for example in the book by Messiah \cite{messiah}, and the radial functions are often written as $\psi_1 = G$ and $\psi_2 = F,$ as in the book by Greiner \cite{greiner}.  We shall assume that the potential $V$ is such that there is a discrete eigenvalue $E$ and that Eqs.(\ref{eq3},~\ref{eq4}) are the eigen-equations for the corresponding radial eigen-states.  Our geometrical method (to be outlined below) will also presume that the potential can be written as a smooth transformation $V(r) = g(-1/r)$, where $g$ is monotonically increasing and of definite convexity. In this paper we shall present the problem explicitly for the cases $d > 1.$ We shall label the discrete eigenvalues by $k_d$ and the number $\nu = \nu_1$ of nodes in the large radial component $\psi_1.$ This convenient labelling is suggested by a result of Rose and Newton \cite{rose} to the effect that if $\nu_2$ is the number of nodes in the small component $\psi_2,$ then if $\tau=-1$, $\nu_2=\nu_1$; and if $\tau=1$, then $\nu_2=\nu_1+1$.
We study the discrete Dirac spectrum generated by the potential $V(r) = V_q(r) = v f_q(r),$ where $v>0$ is the coupling parameter, and the potential shape $f_q(r)$ is given by
\begin{equation*}\label{pot}
f_q(r) =
-\frac{1}{(r^q+b^q)^{\frac{1}{q}}}.
\end{equation*}
This potential represents a family of softcore (truncated) Coulomb potentials,
 which are useful as model potentials in atomic and molecular physics.
In the limit as $q\rightarrow \infty$, the potential descends to the cut-off Coulomb potential $f_{\infty}$ given by
\[
\lim_{q\rightarrow\infty} f_q(r) = f_{\infty}(r) = \left\{
\begin{array}{l l}
-\frac{1}{b},  &{\rm if }\, r < b;\\
\\
-\frac{1}{r},  &{\rm if }\, r \ge b.
\end{array}
\right.
\]

 The bound states are
obtained in terms of three potential parameters: the coupling
$v>0,$ the cut-off parameter $b >0,$ and the power parameter
$q \ge 1.$  The cases $q=1$ and $2$ are of special physical
significance \cite{MP,P1,SVD,DMVD,SR,FF0,CM,OM,MO}. The potential
$f_1$ represents the potential due to a smeared charge and is
useful in describing mesonic atoms. The potential $f_2$ is similar
to the shape of the potential due to a finite nucleus and
experienced by the muon in a muonic atom. Extensive applications
of the softcore Coulomb potential, $f_2$, have been made through
model calculations corresponding to the interaction of intense
laser fields with atoms \cite{LM,JHE1,JHE3,PLK,CWC,CK}. The
parameter $b$ can be related to the strength of the laser
field, with the range $b=20-40$ covering the experimental
laser field strengths \cite{LM}.\medskip

In the non-relativistic case, Mehta and Patil \cite{MP} have presented
analytical solutions for the $s$-state eigenvalues corresponding
to the $f_1$ potential. Also upper and lower energy bounds and some exact analytic solutions
 have been found for special cases with the potentials $f_1$ and $f_2$ \cite{h90,h130}.
Patil \cite{P1} has discussed the analyticity of the scattering phase shifts for two particles interacting through the potentials
 $f_q$ with $q=1$ and $q=2$.
Much less is known concerning
 the corresponding relativistic problem.
\medskip

The principal idea that  is  used in this paper is that of envelope theory.  We suppose that a given potential shape $f(r)$ can be represented as
a smooth transformation $f(r) = g(h(r)),$ where $h(r)$ is a potential that generates a soluble spectral problem.
Since tangents to $g(h)$ are of the form $ah(r) +c,$ they generate a family of soluble `tangential problems'. If the transformation function $g$ has definite convexity, these tangential problems lead via comparison theorems to a set of energy bounds. Envelope theory picks out the best of these.  Although the principal focus of the present paper is on spatial dimensions $d >1,$ we note that in $d = 1$ dimension, the corresponding potential would be of the form $V(x) = v f(|x|),$ and the spatial components $\psi_1(x)$ and $\psi_2(x)$ of the spinor can be   classified as even or odd functions. These components need not now have to vanish at the origin unless the potential is sufficiently singular there. Meanwhile, the normalization would be given by $\int_{-\infty}^{\infty}(\psi_1^2(x) + \psi_2^2(x))dx = 1.$ The geometrical reasoning used in the present paper applies equally well to problems in one dimension but, of course, necessitates useful exactly soluble problems to be used as bases for the approximations; the Dirac Coulomb problem, for example, is problematic in $d=1$ dimension \cite{dong,katsura}. Nieto \cite{nieto} has presented an analysis of non-relativistic problems in $d$ dimensions, with $d$ real and positive. For the future, the task of discussing a similar generalization presents itself for relativistic problems, and smooth transformations thereof.
\medskip

In section 2 we review some recent comparison theorems for Dirac eigenvalues, and we discuss general scaling and  monotonicity properties, In section 3 we describe envelope theory for the Dirac equation, and  in section 4 we  look at the specific case in which the softcore Coulomb potential is written as a smooth convex transformation of a pure Coulomb potential: this  generates a simple formula for lower energy bounds.  We find some of these bounds explicitly and compare them with accurate values found by direct numerical methods.

\section{Comparisons, monotonicity, and scaling}
It is not a simple matter to characterize the discrete Dirac spectrum variationally \cite{franklin,goldman,grant}. However, in spite of this, some comparison theorems have recently been proved \cite{hallds, halldscom,semay}, and we state two of these theorems here for use in the present paper. 

\medskip
\noindent{\bf  Theorem~1} \cite{hallds}~~{\it The real attractive central potential $V(r,a)$ depends smoothly on the parameter $a$, and $E(a)= E_{k_d\nu}(a)$ is a corresponding discrete Dirac eigenvalue. Then:}
\begin{equation*}\label{theorem1}
\partial V/\partial a \geq 0~~\Rightarrow~~E'(a) \geq 0~~~{\it and}~~~\partial V/\partial a \leq 0~~\Rightarrow~~E'(a) \leq 0.
\end{equation*} 

\medskip
\noindent{\bf  Theorem~2} \cite{halldscom}~~{\it Suppose that $E^{(1)}_{k_d\nu}$ and $E^{(2)}_{k_d\nu}$ are Dirac eigenvalues corresponding to two distinct 
attractive central potentials $V^{(1)}(r)$ and $V^{(2)}(r)$.  Then: }
\begin{equation*}\label{theorem2}
V^{(1)}(r) \leq V^{(2)}(r) ~~~ \Rightarrow ~~~E^{(1)}_{k_d\nu} \leq E^{(2)}_{k_d\nu}.
\end{equation*}

If the exact eigenvalues of $H$ are written $E(v,b,q,m)$, then we conclude from Theorem~1 and the monotone behavior of the potential $V(r) = vf_q(r)$ with respect to the parameters that these spectral functions are monotone in each parameter, decreasing in $v$ and $q$, and increasing in $b$. Because
\begin{equation*}\label{monotonicity}
\frac{\partial E}{\partial v} < 0,\quad \frac{\partial E}{\partial b} > 0,
\quad {\rm and}\quad
\frac{\partial E}{\partial q} < 0.
\end{equation*}
We now change variable $r\rightarrow\delta r$ in Eqs.(\ref{eq3},~\ref{eq4}), where $\delta>0$ is constant, multiply through by $\delta$, and compare eigenvalues, we obtain the general scaling law for the family of softcore Coulomb potentials $V(r) = v f_q(r)$ under the Dirac coupled equations, namely
\begin{equation*}\label{dscale}
E(v,b,q,m)=\frac{1}{\delta}E\left(v,\frac{b}{\delta},q,\delta m\right).
\end{equation*}
Choosing $\delta=b$ and $\delta=1/m$ we get, respectively, the special scaling laws
\begin{equation*}\label{dscale2}
E(v,b,q,m)=\frac{1}{b}E\left(v,1,q,b m\right)
=mE\left(v,b m,q,1\right).
\end{equation*}
This much is known in general for the whole class of problems.\medskip

\section{Envelope theory}
Envelope theory has been used since 1980 as a geometrical method of spectral approximation \cite{env1,env2,env3,env4,env6}.  We include here a brief self-contained summary of what is needed for the  present task. 
We consider a potential shape $f(r)$ that can be written as a smooth transformation $f(r) = g(h(r))$ of
a potential $h(r)$ for which the solutions of the Dirac equation are exactly known. In our specific application, $h(r)$ will be chosen as the Coulomb potential $h(r) = -1/r$, and $f(r) = f_q(r)$  is the softcore Coulomb potential, which we are studying.   However, it is clearer to discuss the method  in general at first.  Let us suppose that the transformation function $g(h)$ is monotonically increasing and convex, that is to say, $g'(h) > 0$ and $g''(h)> 0.$ This means that $g(h)$ lies above its tangents.  If $r = t$ is the point of contact between curve and tangent, we have (in the convex case)
a family of `lower' tangential potentials given by
\begin{equation}\label{ft}
f(r) \ge f^{(t)}(r) = a(t)h(r) + c(t),
\end{equation}
where
\begin{equation}\label{atbt}
a(t) = g'(h(t)) \quad {\rm and} \quad c(t) = g(h(t))-h(t)g'(h(t)).
\end{equation}
We note parenthetically that if $g(h)$ is {\it concave}, we obtain, instead, a family of upper bounds. 
Continuing the convex case, if $E=D(u)$ describes how a discrete Dirac eigenvalue corresponding to the potential
$uh(r)$ depends on the coupling $u$, then the potential inequality \eq{ft} and Theorem~2  imply for the original potential $V(r) = vf(r)$ the spectral inequality $E \ge D(va(t))+vc(t).$ This expression can then be optimized over the contact point $t$ to give the lower bound
\begin{equation}\label{et}
E(v) \ge E^{L}(v) = \sup_{t > 0}\left[D(va(t))+vc(t)\right].
\end{equation}
Replacement $a(t)$ and $c(t)$, using \eq{atbt}, in \eq{et} and differentiation with respect to $h$ gives the value for the critical point $h=D'(vg'(h))$. Then we can re-write the right hand side of \eq{et} by changing the minimization variable from $t$ to $u$ by means of the invertible transformation $u = vg'(h(t)),$ yielding the following alternative
 form for the best energy bound:
\begin{equation}\label{eu}
E(v) \ge E^{L}(v) = \sup_{u >0}\left[D(u)-uD'(u) + vg(D'(u))\right].
\end{equation}

\section{Energy bounds for the softcore Coulomb potential}
 We first consider the Dirac equation for the pure Coulomb problem with potential $V(r) = -u/r,$ where the coupling parameter $u=\alpha Z$ is not too large. We write the exact discrete eigenvalues as $D_{k_d\nu}(u) = D(u)$ and they are given \cite{jiang,dong} exactly by 
\begin{eqnarray}
D(u) &=& m\left\{1 + u^{2}\left[\nu - (1 - \tau)/2+ (k_d^{2} - u ^{2})^{\half}\right]^{-2}\right\}^{-\half},\label{eqdirac1}\\
 &=& m\left\{1 + u^{2}\left[n-|k_d|+ (k_d^{2} - u ^{2})^{\half}\right]^{-2}\right\}^{-\half},\,0<u<1,\nonumber
\end{eqnarray}
where 
\begin{equation*}\label{eqtau}
k_d = \tau\left(j + \frac{d-2}{2}\right), \quad \tau=\pm 1,
\end{equation*}
$\nu = 0,1,2, \dots$ is the number of nodes in the upper radial function $\psi_1(r)$, and $n$ is the principal quantum number defined in general 
(for both Coulomb and non--Coulomb central potentials) by 
\begin{equation*}\label{eqpqn}
n =  \nu +|k_d| - \frac{1-\tau}{2}.
\end{equation*}
The spectroscopic designation 
\[
\{s,p,d,\dots\}\leftrightarrow \ell = \{0,1,2,\dots\}
\]
 is then provided by the formula
\begin{equation*}\label{ell}
\ell = |k_d| - \left(\frac{d-1}{2}\right).
\end{equation*}

\begin{figure}
\centering{\includegraphics[height=8cm,width=12cm,angle=-90]{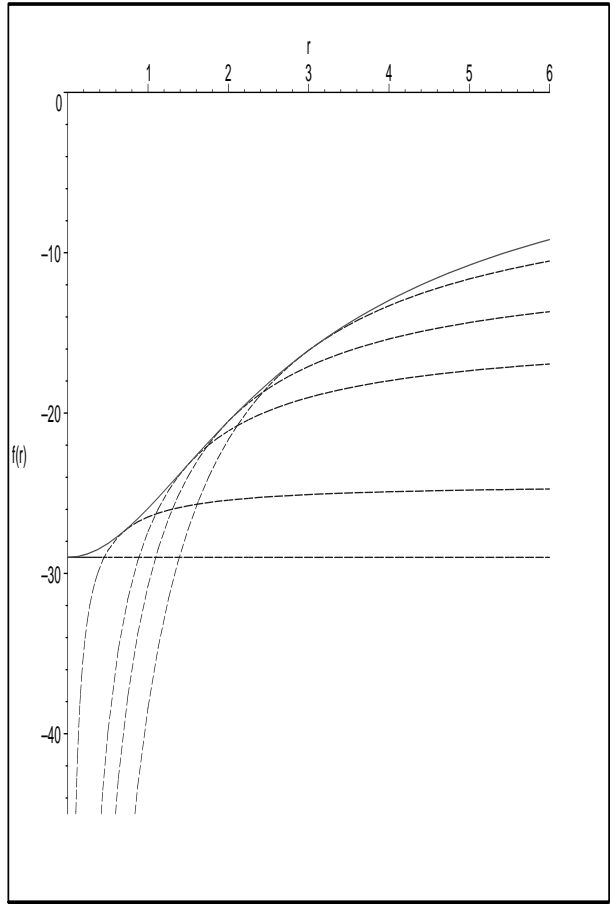}}
\caption{The softcore Coulomb potential for $q = 2$ shown with
 tangential family of shifted Coulomb potentials.}\label{fig1}
\end{figure}

For the softcore Coulomb potential  $f_q(r)$ it follows that the transformation function $g(h)= f_q(-1/h)$ has the explicit form
\begin{equation}\label{gfun}
g(h) = -\frac{1}{\left((-\frac{1}{h})^q +b^q\right)^{\frac{1}{q}}}.
\end{equation}
Consequently $dg/dh$ and $d^2g/dh^2$ are given by
\begin{equation*}\label{gpgpp} 
\frac{dg}{dh}=\frac{\left(-\frac{1}{h}\right)^{q-1}}
{h^2\left[\left(-\frac{1}{h}\right)^q+b^q\right]^{1/q+1}}
\quad {\rm and}\quad 
\frac{d^2g}{dh^2}=\frac{(q+1)b^q\left(-\frac{1}{h}\right)^q}
{h^2\left[\left(-\frac{1}{h}\right)^q+b^q\right]^{1/q+2}}.
\end{equation*}
Since the quantity $-1/h>0$, we conclude that $dg/dh>0$ and $d^2g/dh^2>0$.
Thus $g$ is monotonically increasing and convex; its tangents $f^{(t)}(r)=a(t)h(r)+b(t)$ given by \eq{atbt} are lower bounds to $f_q(r)$. 
The case $q=2$, $b=2$, and $v=58$ is shown in Fig. (\ref{fig1}).

 Lower energy bounds are therefore provided by \eq{eu} with $D(u)$ given by \eq{eqdirac1} and the transformation function $g$ given explicitly by \eq{gfun}. The lower bounds $E^L_{k_d \nu}$ are compared with accurate numerical values $E_{k_d \nu}$ in Table 1. 

\begin{table}
\caption{Comparison the exact softcore Coulomb energy eigenvalues $E$ with the lower bounds $E^L$, for $m=1$.}
\begin{center}
\begin{tabular}{|c|c|c|c|c|c|c|c|c|c|}
\hline
$\nu$ & $v$ & $b$ & $q$ & $d$ & $j$ & $\tau$ & $k_d$ & $E_{k_d \nu}$ & $E^L_{k_d \nu}$\\
\hline
\hline
0&0.9&2&2&2&1/2&-1&-1/2&0.76378&0.69320\\
\hline
0&2&5&5&2&3/2&-1&-3/2&0.72910&0.68877\\
\hline
0&0.1&0.1&1&3&1/2&-1&-1&0.99517&0.99509\\
\hline
0&0.1&0.1&10&3&1/2&-1&-1&0.99499&0.99499\\
\hline
0&0.9&0.5&7&5&3/2&-1&-3&0.95394&0.95394\\
\hline
0&3.1&3&4&10&1/2&-1&-9/2&0.75265&0.74013\\
\hline
1&0.9&2&8&2&7/2&-1&-7/2&0.97956&0.97956\\
\hline
1&0.6&7&6&3&9/2&1&5&0.99631&0.99631\\
\hline
1&0.9&0.2&2&3&1/2&-1&-1&0.88328&0.85076\\
\hline
1&1&0.5&2&6&1/2&-1&-5/2&0.97957&0.95691\\
\hline
1&0.9&0.9&3&8&3/2&1&9/2&0.99028&0.99028\\
\hline
1&0.9&0.1&5&9&3/2&-1&-5&0.98869&0.98863\\
\hline
2&1.1&0.5&1&2&1/2&1&1/2&0.95345&0.75185\\
\hline
2&0.5&8&6&2&5/2&-1&-5/2&0.99417&0.99375\\
\hline
2&0.8&1.5&7&2&7/2&1&7/2&0.99252&0.99230\\
\hline
2&0.6&7&6&3&9/2&-1&-5&0.99632&0.99631\\
\hline
2&2&9&1&4&3/2&-1&-5/2&0.96327&0.95513\\
\hline
2&0.9&0.1&5&6&1/2&1&5/2&0.98605&0.98605\\
\hline
2&1&8&3&9&3/2&1&5&0.99217&0.99209\\
\hline
3&0.7&5&4&3&1/2&1&1&0.99100&0.98916\\
\hline
3&0.9&3&8&6&5/2&-1&-9/2&0.99271&0.99270\\
\hline
4&0.7&5&4&3&1/2&-1&-1&0.99218&0.98916\\
\hline
4&0.9&0.4&4&2&3/2&1&3/2&0.98965&0.98963\\
\hline
4&0.4&6&2&5&3/2&-1&-3&0.99842&0.99836\\
\hline
5&1.1&10&2&4&1/2&-1&-3/2&0.99013&0.98526\\
\hline
6&0.7&0.3&6&3&1/2&-1&-1&0.99685&0.99461\\
\hline
7&0.7&4&2&7&3/2&-1&-4&0.99799&0.99769\\
\hline
8&1.1&9&1&2&5/2&1&5/2&0.99655&0.99589\\
\hline
9&4&3&7&3&1/2&-1&-1&0.99702&0.99655\\
\hline
10&0.7&0.5&5&3&1/2&-1&-1&0.99889&0.99787\\
\hline
\end{tabular}
\end{center}
\end{table}

The eigenvalue formula \eq{eqdirac1} is based on the Coulomb spectral function $D(u)$ which admits couplings $u$ satisfying $u<1$. Thus we require in \eq{et}, that $va(t)<1$ and consequently we cannot consider arbitrarily large coupling $v$ for the softcore Coulomb potential. The Coulomb degeneracy $E_{k_d \nu}= E_{-k_d \nu+1}$ is expressed by \eq{eqdirac1} and, of course, this symmetry is satisfied by the lower bounds. For instance, from Table 1 we see that the lower bounds $E^L_{k_d \nu}$ are degenerate for the pairs $\{E_{1_3 3},\, E_{-1_3 4}\}$ and $\{E_{5_3 1},\, E_{-5_3 2}\}$, although the eigenvalues themselves are not exactly equal.  Nevertheless, the simple lower bound formula \eq{eu} is valid for all the discrete Dirac eigenvalues and is often very effective.

\section{Conclusion}
This paper is based on two  ideas: (1) a comparison theorem valid for a discrete Dirac spectrum, and (2) 
a geometrical theory that generates a family of potentials tangential  to a smooth transformation $g(h)$ of a Dirac-soluble base potential $h.$  These two strands are connected if $g(h)$ has definite convexity  so that its graph lies either above or below its tangents. Meanwhile the tangential potentials are of the form $ah(r)+c$ and Dirac's equation is soluble exactly for each of them. For given values of the potential parameters, and good quantum numbers, one is then able to find the best tangential potential in the sense of providing the best energy bound. We have used the Coulomb envelope base $h(r) = - 1/r$ to generate optimized energy lower bounds for a family of softcore Coulomb potentials given by $g(h(r)) = -1/{(r^q+b^q)^{\frac{1}{q}}}$, where $b>0$ and $q \ge 1$.  
Because the potential is central, the geometric argument leading to the lower bound via the Dirac comparison theorem transcends the question of the number $d$ of spatial dimensions. The estimate is an energy bound whenever $g$ has definite convexity. 

\section*{Acknowledgements}

One of us (RLH) gratefully acknowledges partial financial support
of this research under Grant No.\ GP3438 from the Natural Sciences
and Engineering Research Council of Canada.

\medskip


\section*{References}
\begin{thebibliography}{99}

\bibitem{jiang}Y. Jiang, J. Phys. A {\bf 38} 1157 (2005).
\bibitem{messiah}A. Messiah, {\it Quantum Mechanics}, (North Holland, Amsterdam, 1962). The Dirac equation for central fields is discussed on page 928.
\bibitem{greiner}W. Greiner, {\it Relativistic Quantum Mechanics}, (Springer, Heidelberg, 1990). The Dirac equation for the Coulomb central potential is discussed on page 178.
\bibitem{rose}M.~E.~Rose and R.~R.~Newton, Phys. Rev. {\bf 82}, 470 (1951).
\bibitem{MP}C. H. Mehta and S. H. Patil , Phys. Rev. A \textbf{17}, 43 (1978).
\bibitem{P1}S. H. Patil, Phys. Rev. A \textbf{24}, 2913 (1981).
\bibitem{SVD}D. Singh, Y. P. Varshni, and R. Dutt, Phys. Rev. A \textbf{32}, 619 (1985).
\bibitem{DMVD}H. De Meyer and G. Vanden Berghe, J. Phys. A: Math. Gen. \textbf{23}, 1323 (1990).
\bibitem{SR}A. Sinha and R. Roychoudhury, J. Phys. A: Math. Gen. \textbf{23}, 3869 (1990).
\bibitem{FF0}F. M. Fern\'andez, J. Phys. A: Math. Gen. \textbf{24}, 1351 (1991).
\bibitem{CM}R. N. Chaudhuri and M. Mondal, Pramana-J.Phys., \textbf{39}, 493 (1992).
\bibitem{OM}M. Odeh and O. Mustafa, J. Phys. A: Math. Gen. \textbf{33}, 7013 (2000).
\bibitem{MO} O. Mustafa and M. Odeh, J. Phys. B: At. Mol. Opt. Phys. \textbf{32}, 3055 (1999).
\bibitem{LM}C. A. S. Lima and L. C. M. Miranda, Phys. Rev. A \textbf{23}, 3335 (1981).
\bibitem{JHE1} J. H. Eberly, Q. Su, and J. Javanainen, Phys. Rev. Lett. \textbf{62}, 881 (1989).
\bibitem{JHE3} Q. Su and J. H. Eberly, Phys. Rev. A \textbf{44}, 5997 (1991).
\bibitem{PLK} M. Protopapas, C. H. Keitel, and P. L. Knight, Rep. Prog. Phys. \textbf{60}, 389
(1997). References therein.
\bibitem{CWC} C. W. Clark, J. Phys. B \textbf{30}, 2517 (1997).
\bibitem{CK} Y. I. Salamin, S. H. Hu, K. Z. Hatsagortsyan, C. H. Keitel, Phys. Rev. \textbf{427}, 41 (2006).
\bibitem{h90} R.~L.~Hall, and Q. D. Katatbeh 
Phys. Lett. A {\bf 294}, 163-167 (2002).
\bibitem{h130} R.~L.~Hall, N. Saad, K. D. Sen, and H. Ciftci 
Phys. Rev. A {\bf 80}, 032507 (2009).
\bibitem{dong}S.~H.~Dong, J. Phys. A {\bf 36}, 4977 (2003).
\bibitem{katsura}H. Katsura and H. Aoki, J. Math. Phys. {\bf 47}, 032301 (2006).
\bibitem{nieto}M. M. Nieto, Phys. Lett. A {\bf 293}, 10 (2002).
\bibitem{franklin}J.~Franklin and R.~L.~Intemann, Phys. Rev. Lett. {\bf 54}, 2068 (1985).
\bibitem{goldman}S.~P.~Goldman, Phys. Rev. A {\bf 31}, 3541 (1985).
\bibitem{grant}I.~P.~Grant and H.~M.~Quiney, Phys. Rev. A {\bf 62}, 022508 (2000).
\bibitem{hallds}R.~L.~Hall, Phys. Rev. Lett. {\bf 101},  090401 (2008).
\bibitem{halldscom}R.~L.~Hall, Phys. Rev. A {\bf 81},  052101 (2010).
\bibitem{semay}C. Semay, Phys. Rev. A {\bf83}, 024101 (2011).
\bibitem{env1}R.~L.~Hall, Phys. Rev. D {\bf 22}, 2062 (1980).
\bibitem{env2}R.~L.~Hall, J. Math. Phys. {\bf 24}, 324 (1983).
\bibitem{env3}R.~L.~Hall, J. Math. Phys. {\bf 25}, 2708 (1984).
\bibitem{env4}R.~L.~Hall, Phys. Rev. A {\bf 39}, 550 (1989).
\bibitem{env6}R.~L.~Hall, J. Math. Phys. {\bf 34}, 2779 (1993).



\end{thebibliography}
\end{document}